\documentclass[twocolumn,showpacs,preprintnumbers,amsmath,amssymb]{revtex4}

\usepackage{graphicx}
\usepackage{dcolumn}
\usepackage{bm}
\usepackage{color}

\begin{document}

\preprint{xxx}

\title{A Xylophone Configuration for a third Generation Gravitational Wave Detector}

\author{Stefan Hild}
 \email{hild@star.sr.bham.ac.uk}
\affiliation{School of Physics and Astronomy, University of
Birmingham, Edgbaston, Birmingham B15 2TT, UK}
\author{Simon Chelkowski}
\affiliation{School of Physics and Astronomy, University of
Birmingham, Edgbaston, Birmingham B15 2TT, UK}
\author{Andreas Freise}
\affiliation{School of Physics and Astronomy, University of
Birmingham, Edgbaston, Birmingham B15 2TT, UK}
\author{Janyce Franc}
\affiliation{Laboratoire des Materiaux Avances (LMA), 22 Boulevard Niels Bohr, 
Villeurbanne Cedex,  69622, France }
\author{Nazario Morgado}
\affiliation{Laboratoire des Materiaux Avances (LMA), 22 Boulevard Niels Bohr, 
Villeurbanne Cedex,  69622, France }
\author{Raffaele Flaminio}
\affiliation{Laboratoire des Materiaux Avances (LMA), 22 Boulevard Niels Bohr, 
Villeurbanne Cedex,  69622, France }
\author{Riccardo DeSalvo }
\affiliation{California Institute of Technology, LIGO Project, Pasadena, CA
91125, USA}

\date{\today}

\begin{abstract}

Achieving the demanding sensitivity and bandwidth, envisaged
for third generation gravitational wave (GW) observatories, is extremely  challenging
 with a single broadband interferometer. 
Very high optical powers
(Megawatts) are required to reduce the quantum noise contribution at 
high frequencies, while  the interferometer mirrors have to be cooled
to cryogenic temperatures in order to reduce thermal noise sources at low frequencies.
To resolve this potential conflict of cryogenic test masses with high thermal load,
  we present a conceptual design for a 2-band xylophone configuration for a third 
generation GW observatory, composed of a high-power, high-frequency interferometer
and a cryogenic low-power, low-frequency instrument. Featuring inspiral ranges of
 3200\,Mpc and 38000\,Mpc
for binary neutron stars and binary black holes coalesences, respectively, we 
find that the potential sensitivity of xylophone configurations can be significantly wider and
better than what is possible in a single 
 broadband interferometer.
\end{abstract}


\maketitle

\section{\label{sec:Intro}Introduction}

Over the last decades scientists pioneered the field of laser-interferometric 
gravitational wave (GW) detection, culminating in the establishment of a 
worldwide network of large-scale gravitational wave detectors \cite{ligo, virgo, geo}. 
 
The design and
construction of  a
 second generation of GW observatories is well underway and observation 
with  ten times improved sensitivity is expected to start in about 5 years 
\cite{aligo2, adv, Willke06}.
Triggered by the Einstein GW Telescope (ET) design study within the European
 FP7 framework \cite{ET}, research 
has started on design options for a third generation GW observatory \cite{Freise09, 
Hild08}, aiming for a sensitivity 100 times better than of current instruments and thus
allowing us to scan a one million times larger fraction of the Universe for
 astrophysical GW sources. In addition to improved sensitivity, a key feature 
of observatories such as ET will be their strongly expanded bandwidth, covering the
range from 1\,Hz to 10\,kHz.
  Especially the extension of the detection band towards the lower 
frequency end will increase the number and signal-to-noise ratio of observable gravitational
wave signals and 
therefore significantly enhance the astrophysical impact of third generation 
observatories \cite{sathya}. 

As we will show in Section \ref{sec:Benefit}, for achieving the immense bandwidth
envisaged for instruments such as ET, it might be highly beneficial, if not even 
technically unavoidable
 to split the detection band into several optimized  detectors of moderate bandwidth,
forming altogether a so-called \textit{Xylophone} interferometer covering the
 full detection band. In Section \ref{sec:Example} we present for the first time a 
potential design for a third generation xylophone configuration, consisting of a low-power, 
cryogenic interferometer optimized for the low-frequency band and a higher-power,
room-temperature interferometer covering the high-frequency band.

\begin{figure*}[htb]
\centerline{
\includegraphics[width=0.45\textwidth,keepaspectratio]{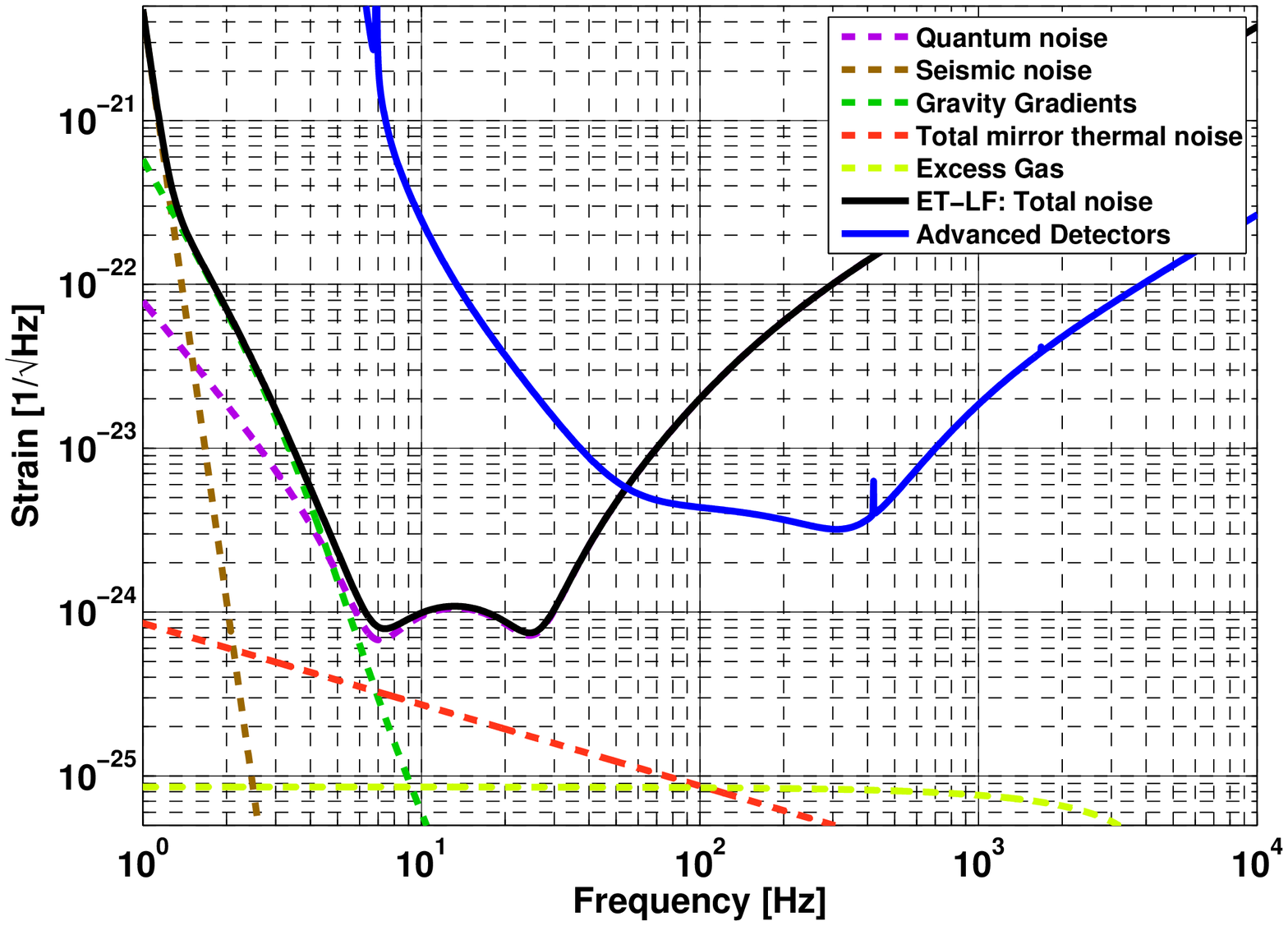}
\includegraphics[width=0.45\textwidth,keepaspectratio]{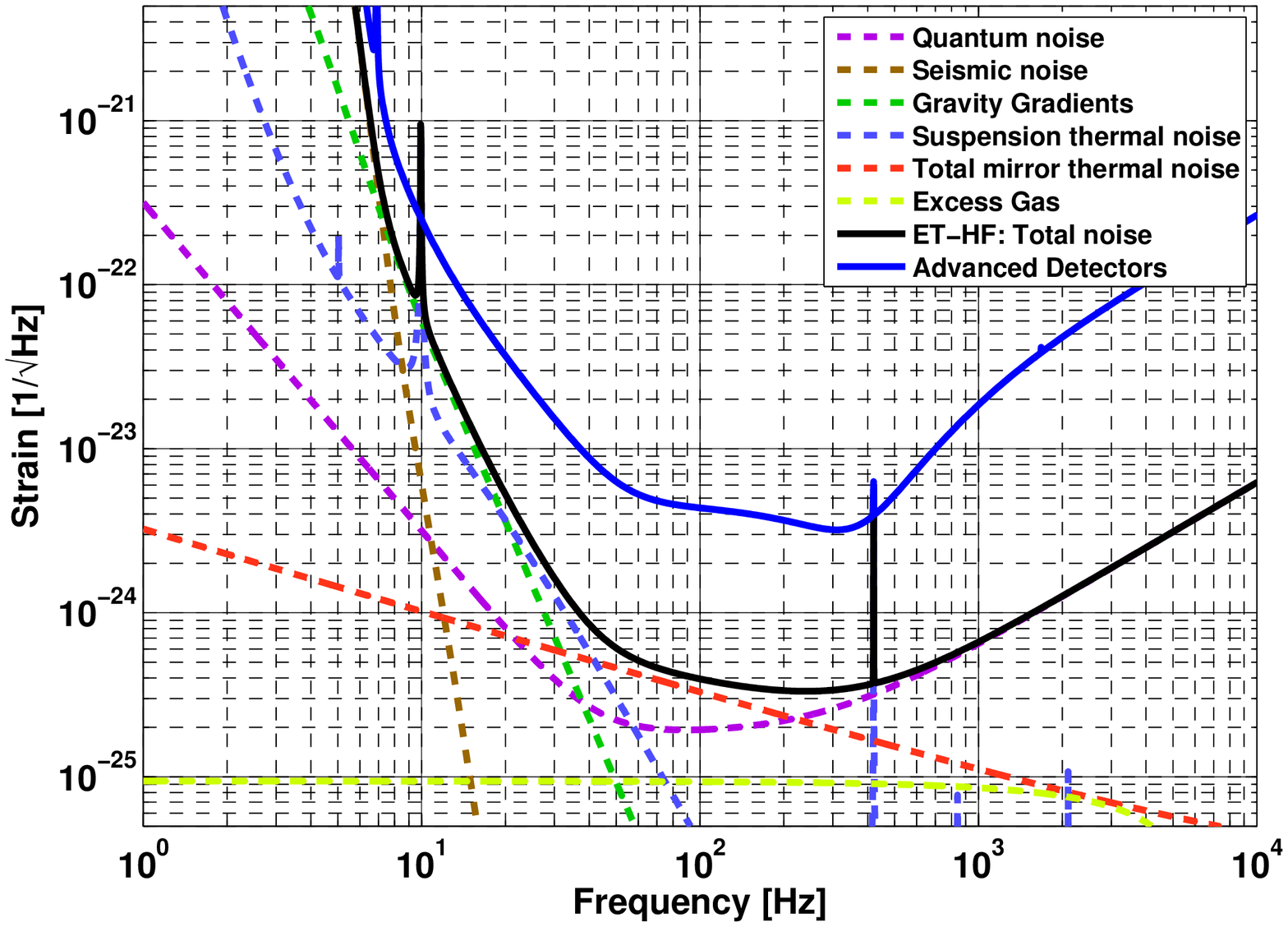}
} \caption{Fundamental noise contributions of the two analysed xylophone detectors.
While the high frequency interferometer, ET-HF, (right plot) features a circulating light power of 
3\,MW, the low-frequency detector, ET-LF, (left plot) operates with only 18\,kW of circulating power, 
allowing easier use of cryogenic test masses. The main parameters of both interferometers
are compared in Table \ref{tab:summary}.  As a comparison, the sensitivity curve of the 'Advanced'
 detectors is shown in blue.
} \label{fig:noise_budget}
\end{figure*}

\section{\label{sec:Benefit}Potential Benefits of Xylophone configurations
for third generation gravitational wave detectors}

Spanning the detection band over four orders of magnitude in frequency, 
as it is ask for third generation GW observatories such as ET,
is technically extremely challenging: Different noise types dominate the various 
frequency bands and often show opposite response for different
tuning of the same design parameter.

 A well-known example for such a behavior is the
correlation of the two quantum noise components: photon shot noise (PSN) and 
photon radiation pressure noise (PRPN). In order
to improve the PSN limited sensitivity at  high frequencies one needs to increase 
circulating optical power of the GW detector, which at the same increases the 
PRPN and therefore worsens the low frequency sensitivity. 
Vice versa, lowering the circulating power reduces the PRPN
and improves the low frequency sensitivity, while the PSN contribution will raise
and reduce the high frequency sensitivity. 

This dilemma can be resolved by following the path of
 electromagnetic astronomy, where telescopes are built for
a specific, rather narrow-banded detection window (visible, infrared etc) and
later on the data from different frequency bands is combined to cover the 
desired bandwidth.  Building two or more GW detectors, each optimised for reducing
the noise sources at one specific frequency band, 
can form a xylophone observatory providing substantially improved broadband sensitivity.

The xylophone concept was first suggested for Advanced LIGO,
 proposing to complement the standard broadband 
 interferometers with an interferometer optimized for lower frequency,
 thus enhancing the detection of high-mass binary systems
\cite{Shoemaker01, DeSalvo04, conforto03}. The concept was then taken forward
 for underground observatories \cite{Aspen}. 
 In this article we extend the xylophone concept
 for the application in third generation GW observatories.

One may think that a xylophone might significantly increase the required
hardware and its cost, i.e. building more than one
 broadband instrument. However, such an argument
 does not take the technical 
simplifications that it would allow, the better reliability of simpler instruments,
 and the more extensive scientific reach allowable into account. 
For example splitting a third generation observatory in to
a low-power, low-frequency  and a high-power high-frequency
interferometer, has not only the potential to resolve the above mentioned 
conflict of PSN and PRPN, but
also allows to avoid the combination of high optical power and cryogenic test masses.
To reduce thermal noise to an acceptable level in the low frequency band, it is
expected that cryogenic suspensions and test masses are required. 
Even though tiny, the residual absorption of the
dielectric mirror coatings deposits a significant amount of heat
 in the mirrors, which is difficult to extract,
without spoiling the performance of the seismic isolation systems, and thus 
limiting the maximum circulating power of a cryogenic interferometer.

\section{\label{sec:Example} Example of a 2-band xylophone  configuration
for the Einstein GW Telescope (ET)}

 Starting from the single-detector ET configuration
described in \cite{Hild08} we developed a 2-band xylophone detector configuration
to resolve the high-power low-temperature problem of a single band ET
observatory. Table \ref{tab:summary} gives a brief overview of the main parameters
of the analysed low-frequency (ET-LF) and high-frequency (ET-HF) detector.

\subsection{ET-HF detector}
The high-frequency interferometer, ET-HF, is an up-scaled but otherwise 
only moderately advanced version of a second
generation interferometer: We considered an arm length of 10\,km 
 and a circulating light power of 3\,MW. 
In order to achieve the aimed high frequency sensitivity we also assumed the 
implementation of squeezed light \cite{Chelkowski05} as well as
tuned signal recycling (SR) \cite{Hild07a}, which allows to simultaneously extract
 both signal sidebands.

To reduce the thermal noise contributions, limiting the medium frequency
range, without recurring to cryogenic temperatures we considered  increasing the 
beam size to the technical maximal feasible value of about 12\,cm beam radius,
as well as changing the beam shape from the currently used TEM$_{00}$
to mesa beams \cite{Galdi06} or a higher order
 Laguerre Gauss (LG) mode \cite{Mours06, Chelkowski09}. Using the
 LG$_{33}$ mode the coating Brownian and the substrate Brownian noise are reduced
by factors 1.61 and 1.40, respectively \cite{Vinet}. 
Please note that the suspension
system of ET-HF is  identical to a second generation GW observatory, but scaled up
 to cope with the higher mirror mass of 200\,kg,
 required to manage
the larger beams with a feasible mirror aspect ratio \cite{Somiya09}.   
The sensitivity curve and the noise budget of ET-HF
is shown in the right hand plot of Figure \ref{fig:noise_budget}.

\begin{table}
\begin{center}
\begin{tabular}{|l|c|c|}
\hline 
Parameter & ET-HF   & ET-LF \\
\hline
Arm length & 10\,km & 10\,km \\
Input power (after IMC) & 500\,W & 3\,W \\
Arm power & ~3\,MW & ~18\,kW\\
Temperature & 290\,K &  10\,K  \\
Mirror material & Fused Silica & Silicon \\ 
Mirror diameter / thickness & 62\,cm / 30\,cm & 62\,cm / 30\,cm \\
Mirror masses & 200\,kg & 211\,kg \\
Laser wavelength & 1064\,nm & 1550\,nm \\
SR-phase & tuned (0.0) & detuned (0.6)\\
SR transmittance & 10\,\% & 20\,\% \\
Quantum noise suppression & 10\,dB & 10\,dB \\
Beam shape &  LG$_{33}$& TEM$_{00}$\\
Beam radius & 7.25\,cm & 12\,cm \\
Clipping loss & 1.6\,ppm & 1.6\,ppm \\
Suspension & Superattenuator & $5 \times 10\,$m \\
Seismic (for $f>1$\,Hz) & $1\cdot 10^{-7}\,{\rm m}/f^2$  & $5\cdot 10^{-9}\,{\rm m}/f^2$  \\
Gravity gradient subtraction & none & factor 50 \\
\hline
\end{tabular}
\caption{Summary of the most important parameters of the 2-band
xylophone detector shown in Figure \ref{fig:noise_budget} . \label{tab:summary}}
\end{center}
\end{table}

\subsection{ET-LF detector}
Unlike ET-HF the low frequency xylophone interferometer, ET-LF, will
require several innovative techniques, well beyond the
scope of first and second generation GW interferometers. In order to
reduce seismic noise, we assumed  an extremely long suspension system, 
composed of 5 stages, each 10\,m tall, in addition to the reduced
seismic level of an underground location \cite{Ohasi03}.
 Even though the reduced seismic excitation
 of an underground site decreases the gravity gradient noise significantly, a further 
 reduction of a factor 50 is required 
 from subtraction of gravity gradient noise. 

\begin{figure}[htb]
\centerline{
\includegraphics[width=0.48\textwidth,keepaspectratio]{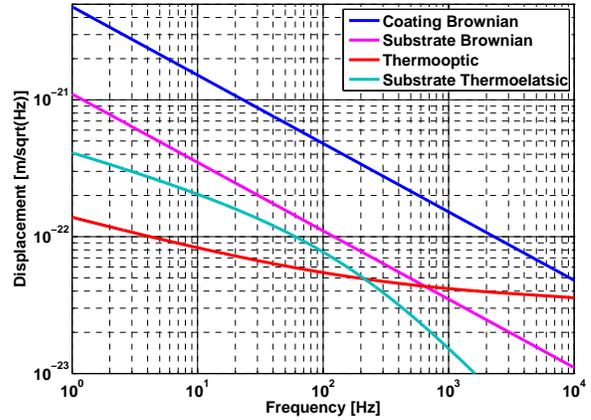}
} \caption{Apparent displacement from individual thermal noise contributions 
for a single cryogenic silicon test mass of ET-LF. 
} \label{fig:single_cryo_mass}
\end{figure}

The main feature of the LF detector is that all thermal noise sources are 
significantly reduced by using cryogenic test masses, which is made possible 
by the reduced 
 optical power of only  18\,kW,   comparable to that of a  first
generation GW detector. Sapphire \cite{Uchiyama99} and silicon have 
been  proposed as test mass material for a cryogenic GW detector. 
However, material costs and material properties, as well as the 
available boule dimensions \footnote{It seems for sure that in a few years
silicon boules with a diameter of 45\,cm will be available. It is not clear whether by the 
time of ET construction the here considered diameter of 62\,cm  will be available.
 However, even in case ET-LF would be constructed with
silicon test masses of only 45\,cm diameter, the coating Brownian noise
would only be increased by a factor $62/45 = 1.38$, which would not
significantly change the sensitivity of ET-LF (see Figure \ref{fig:noise_budget})} 
seem to slightly favor silicon. Therefore, 
 we considered silicon test masses cooled to a temperature 
of 10\,K in this article. The most important material
parameters used in our analysis are the Youngs modulus of 10\,K Silicon of 
164\,GPa and the loss angles of $5 \cdot 10^{-5}$ and $2 \cdot 10^{-4}$ 
for the low and high refraction coating materials, respectively. 

Unfortunately the available measurements indicate  higher
loss angles for the coating materials at cryogenic temperatures than at room
temperature \cite{Martin08}. However, since research on cryogenic coatings just started,
  we optimistically assumed \footnote{Please note that even in case the loss angles
of cryogenic coatings cannot be improved in future, the total mirror thermal noise trace 
in Figure \ref{fig:noise_budget}  would only increase by about a factor of 2, yielding
only a very minor decrease of the ET-LF sensitivity.}  that by
the time construction of third generation instruments starts, coatings will be 
available featuring the same loss angles as current coatings at room temperature
\cite{Harry06, harry07}. The resulting thermal noise contributions of a single cryogenic
silicon test mass are shown in Figure \ref{fig:single_cryo_mass}. 

Using  silicon mirrors also implies to change the laser wavelength
from 1064\,nm to 1550\,nm where Silicon is highly transmissive and
has very low absorption \cite{Green95}. Changing the laser wavelength
has an impact on coating Brownian noise and quantum noise. Due to the fact 
that for 1550\,nm light the mirror coatings have to  be about 1.5 times thicker \footnote{This 
assumes coating materials with the same index of refraction as used for a wavelength of 1064\,nm.},
the overall coating Brownian noise is increased by a factor $\sqrt{(1550/1064)} = 
1.2$. In addition the PSN is also increased by a factor 1.2, while the
PRPN is improved by a factor 1.2.

The resulting noise buget of ET-LF,  limited by
gravity gradient noise at low frequencies and quantum noise at all other 
frequencies, is shown in the left hand plot of Figure \ref{fig:noise_budget}.   
Please note
that we omitted suspension thermal noise from our analysis of ET-LF,
as this is subject of ongoing research and so far no mature noise estimate exists. 
 However, it appears likely that
 the low loss characteristics of crystalline fibers at low temperature may not
 be a limitation above the gravity gradient level.

\subsection{Projected sensitivity of the xylophone configuration}

\begin{figure}[htb]
\centerline{
\includegraphics[width=0.5\textwidth,keepaspectratio]{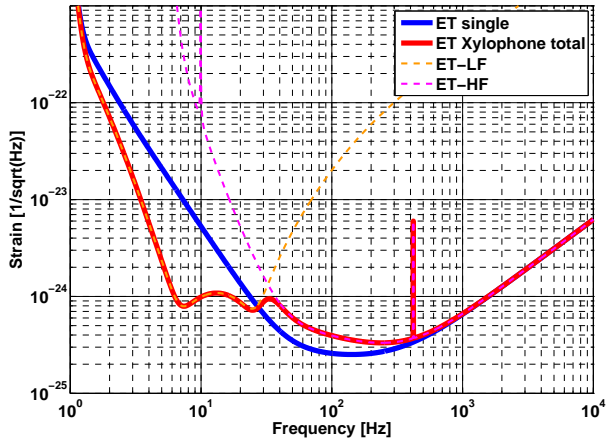}
} \caption{Strain sensitivities of the two presented xylophone detectors (ET-LF and ET-HF)
 and the resulting total instruments sensitivity (ET xylophone total) in comparison to the
sensitivity of the broadband ET interferometer (ET single) from \cite{Hild08}. (Please note that 
no contributions from suspension thermal noise are included for ET single and ET-LF.) 
} \label{fig:h_summary}
\end{figure}

The overall strain sensitivity of the proposed xylophone configuration is shown 
in Figure \ref{fig:h_summary} and compared to the sensitivity of the single 
broadband ET  described in \cite{Hild08}. The resulting 
inspiral ranges  \footnote{We considered NS of 1.4 solar masses 
and BH of 30 solar masses. The inspiral 
ranges are calculated for averaged sky location and a snr of 8.} 
 of the  xylophone are with 3200\,Mpc and 38000\,Mpc
for binary neutron stars (BNS) and binary black holes (BBH), respectively, significantly
larger than the ones for the ET single configuration (BNS range = 2650\,Mpc,
 BBH range = 25000\,Mpc). The sensitivity of the xylophone
 in the intermediate frequency range (50 to 300\,Hz) is 
 slightly worse than the one of ET-single,
 but the overall inspiral ranges improve due to the strongly 
increased sensitivity around 10\,Hz. While the ET-single interferometer is limited by 
RPN between 2 and 30\,Hz, ET-LF can make use of a narrow-band
detuned signal recycling to further decrease the quantum noise.

\section{\label{sec:Summary} Summary and Outlook}

We presented an initial design of a xylophone interferometer for a third generation
GW observatory, composed of a high-power, high-frequency interferometer
complemented by a cryogenic low-power, low-frequency interferometer. The xylophone 
concept  provides a feasible alternative (decoupling
the requirements of high-power laser beams and cryogenic mirror cooling)  compared 
to a single broadband interferometer (ET-single) and is found to potentially
give significantly improved senstivity.

Future efforts
will focus  on investigating the prospects of additional xylophone interferometer 
 either to improve 
the peak sensitivity around 100\,Hz or to push the low frequency wall further down
in frequency.

\begin{acknowledgments}

This work has been supported by the Science and Technology 
Facilities Council (STFC), the European Gravitational Observatory (EGO), the 
Centre national de la recherche scientifique (CNRS), the United States National
Science Foundation (NSF) and the Seventh Framework Programme (Grant 
Agreement 211743) of the European Commission.
\end{acknowledgments}


\newpage

\end{document}